\documentclass[english,american]{elsarticle}
\usepackage{custom-packages}
\usepackage{custom-commands}
\RequirePackage[article]{custom-configuration}

\usepackage{natbib}

\begin{document}
\begin{abstract}
We study an elementary path problem which appears in the pricing step of a column generation scheme solving the \kep.
The latter aims at finding exchanges of donations in a pool of patients and donors of kidney transplantations.
Informally, the problem is to determine \corbis{a} set of cycles and chains of limited length maximizing a medical benefit in a directed graph.
The cycle formulation, a large-scale model of the problem restricted to cycles of donation, is efficiently solved via \bp.
When including chains of donation however, the pricing subproblem becomes \nphard.
This article proposes a new complete \cg scheme that takes into account these chains initiated by altruistic donors.
The development of non-exact dynamic approaches for the pricing problem, the \ngr relaxation and the \cd heuristic, \corbis{leads to an efficient \cg process}. 
%
%
%
\end{abstract}

%
\begin{frontmatter}
  \title{Dealing with elementary paths in the Kidney Exchange Problem}

  \author[myu]{Lucie Pansart\corref{cor1}}
  \ead{lucie.pansart@g-scop.grenoble-inp.fr}

  \author[myu]{Hadrien Cambazard}
  \ead{hadrien.cambazard@g-scop.grenoble-inp.fr}

  \author[myu]{Nicolas Catusse}
  \ead{nicolas.catusse@g-scop.grenoble-inp.fr}


  \address[myu]{Univ.Grenoble Alpes, CNRS, G-SCOP, 38000 Grenoble,
    France}

  \cortext[cor1]{Corresponding author}

\end{frontmatter}

\begin{sloppypar}

\section{Introduction}

\begin{figure}[htbp]
    \centering
        \begin{minipage}{.5\linewidth}
        \begin{subfigure}[b]{\linewidth}
            \centering \includegraphics[page=1 , scale=0.6]{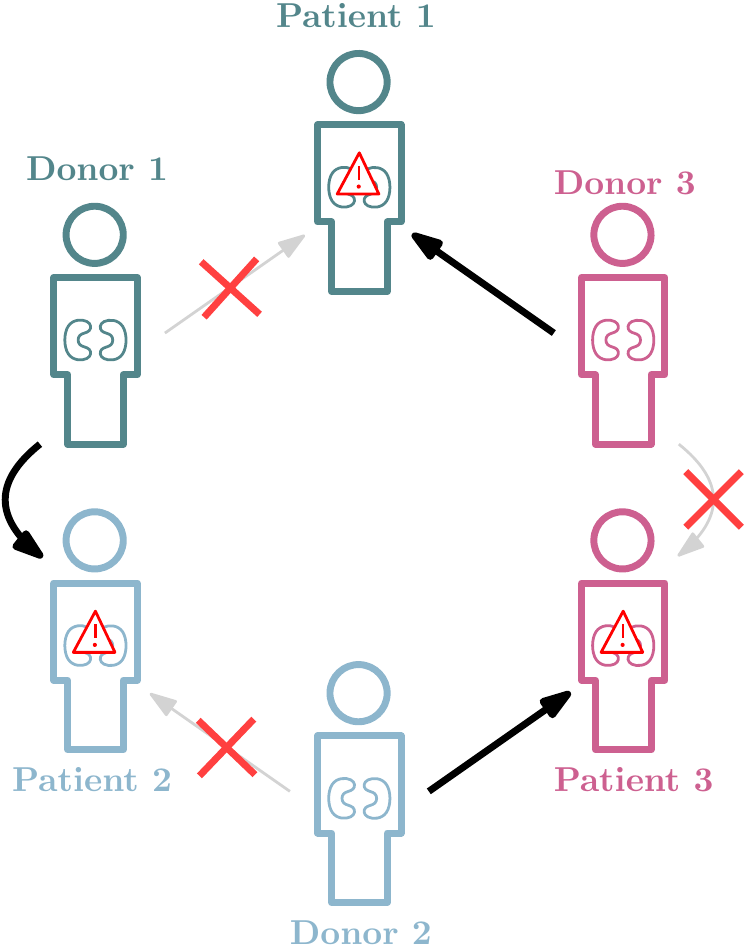}
        	\caption{Cycle of donation with 3 pairs}
        	\label{fig:KEP3}
        \end{subfigure} 
    \end{minipage}
    \begin{minipage}{.4\linewidth}
            \begin{subfigure}[t]{\linewidth}
                \centering \includegraphics[page=1 , scale=0.6]{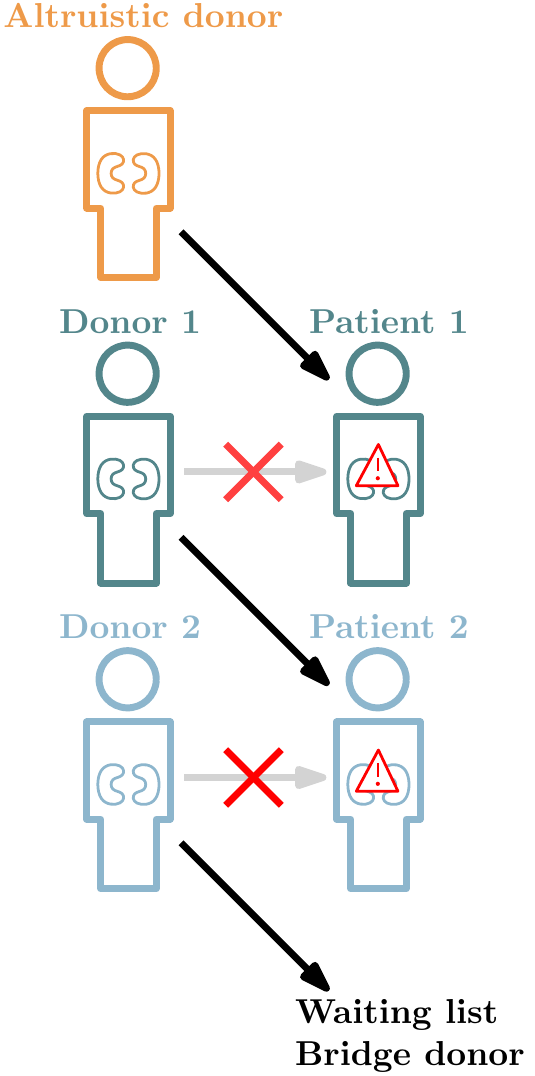}
        	\caption{Chain of donation}
        	\label{fig:KEPc}
            \end{subfigure}
        \end{minipage}
         \captionsetup{justification=centering}
    \caption{Standard exchanges in a \kepg. Chains and cycles may include more pairs.}\label{fig:exchange}
\end{figure}

The \kep models the barter market of \kepgs, which were created to match patients waiting for a kidney transplant to living donors with the objective to find the best possible transplants to perform.  
Each participating patient is paired with a willing, but incompatible, donor.  
This donor accepts to give one kidney only if its associated patient receives one, creating cycles of donation (see Figure \ref{fig:KEP3}). 
It is imperative that a patient is transplanted if its associated donor gives a kidney, so no donors should give before another and all the surgeries of a cycle must be done simultaneously.  
As a cycle requires twice its size operating teams and rooms, \kepgs impose a limit $\k$ on the size of a cycle.
In a lot of countries, programs also include altruistic donors.
These donors are not expecting any transplant to happen in return, creating chains of donation (or domino chains, see Figure \ref{fig:KEPc}).  
In this case, the simultaneity may not be required, as no donor would give his kidney before its patient receives another one.  
Yet, the failure of a transplant causes the failure of every remaining transplant in the chain. 
Thus, it is preferable to have several shorter chains than a big one and  a lot of programs impose a limit $\l$ (usually $\l>\k$) on the length of a chain.
An exchange in a \kepg is therefore either a cycle or a chain of donation which implies at most $\k$ (resp. $\l$) transplants.
The \kep (\skep) aims at finding the best set of exchanges in order to maximize the medical benefit of the performed transplants. 

%
%
%
%
%


The idea of kidney exchange was first mentioned by Rapaport in 1986~\cite{rapaport1986case} and quickly set up in South Korea in 1991~\cite{kwak1999exchange,park2004relay}.
This idea was promising for this country where the public opinion is hostile on deceased transplantation.
The Switzerland was the first European country to perform a kidney exchange in 1999, but the first national \kepg (KPD) in Europe was created by the Netherlands in 2004~\cite{de2005dutch}.
Since then, a dozen states in Europe have created their own KPD~\cite{cost2017kidney}.
In the rest of the world, and to the best of our knowledge, such programs exist only in Canada, Australia and the USA.
We refer the reader to the survey of Ellison~\cite{ellison2014systematic} for more details on the development of KPDs.
It is worth to note that \kepgs involve more and more participants as the usage is spreading in hospitals, but also due to transnational exchanges.
The bigger the pool, the higher the chance to match patients, but the harder the \kep.
\corbis{In 2019}, the largest program in Europe involves 250 \corbis{British} patients~\cite{biro2019building}, but considering that more than half a million Europeans~\cite{kramer2019european} and even more USA citizens~\cite{saran2017us,saran2018us} are treated for end stage kidney disease, new efficient techniques must be developed to handle many more candidates for kidney exchange programs.

Different approaches exist to solve the \kep.
When it contains only cycles of length 2, the \skep can be solved polynomially as a matching problem via Edmonds' algorithm, but as soon as $\k>2$, the problem is  proved to be \npc~\cite{abraham2007clearing,biro2009maximum}.
Consequently, the \skep is often tackled with integer programs and the major ones are surveyed by Mak-Hau~\cite{mak2017kidney}.

We focus on the \textit{cycle formulation}. The original version of this formulation does not take into account altruistic donors and requires to compute every possible cycles.
Abraham \al developed a column generation approach to solve this integer program~\cite{abraham2007clearing}, which is still \corbis{to this day} the best way to solve the \skep without chains of donation.
The cycle formulation can be equivalently applied when including these chains, and we will refer to it as the \textit{\exf}.
In Chen \al, all exchanges are computed beforehand~\cite{chen2011computerized}, but this is not a viable method when the patients pool grows.
On the basis of Abraham \al work, branch-and-price algorithms were developed~\cite{glorie2012iterative,glorie2014kidney,klimentova2014new,plaut2016fast} claiming to accommodate well altruistic donors via chains of donation.
However some of these algorithms (in~\cite{glorie2012iterative,glorie2014kidney,plaut2016fast}) were proven wrong by Plaut \al~\cite{plaut2016hardness} and Klimentova \al did not test their algorithm with altruistic donors~\cite{klimentova2014new}.
Actually, Plaut \al proved in 2016 that the pricing algorithm becomes \npc in this case.

\correction{
\corbis{In order to handle large-scale instances of the \skep, our objective is to address this \nphard problem and establish efficient pricing strategies.}
We study its optimization version, that we refer to as the \empplc (\sempplc).
Based on a review of similar problems, we propose to generate lower and upper bounds using dynamic programming.
\corbis{This approach allows an efficient \cg process and thus to compute the linear relaxation of the \exf.}
This upper bound on the optimal value can be used to assess the quality of the feasible solution that our algorithm constructs by solving the \exf on generated columns.
Our method turns out to be very efficient as the gap between lower and upper bounds is always smaller than 0.5\%, even on instances with more than 800 vertices.
\corbis{The number of patients of these instances is larger than in the current literature or in the field, but is likely to be prevalent in a near future.}

}


In Section~\ref{sec:pf} we model the \kep with the \exf. 
Section~\ref{sec:ELPPRC} defines the \empplc and reviews the different approaches used to solve it, in particular the key idea of our contributions. 
We detail in Sections~\ref{sec:empplc-CD} and~\ref{sec:empplc-NG} the improvements we developed on the \ngr relaxation and the \cd.
\corbis{Their performance are compared in Section~\ref{sec:empplc-exp}.
Finally Section~\ref{sec:CG} shows how these algorithms are used to solve the \kep.}
Note that the path problem studied in this article may be found in other applications and the contributions presented here used in these other cases as well.


\section{Using an exponential formulation for the KEP \label{sec:pf}}
%

\begin{figure}[htbp]
    \centering
    \begin{subfigure}[b]{0.44\textwidth} 
        \centering \includegraphics[page=2 , scale=0.7]{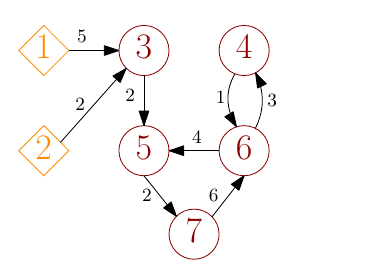}
        \caption{Model}\label{fig:KEPcgraph}
    \end{subfigure}
    ~
    \begin{subfigure}[b]{0.45\textwidth}
        \centering \includegraphics[page=3 , scale=0.7]{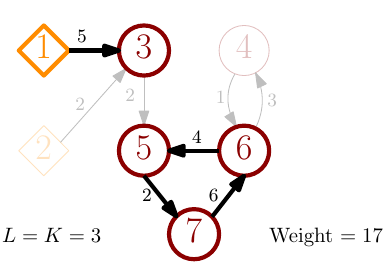}
        \caption{Solution of the \skep}\label{fig:KEPcsol}
    \end{subfigure}
  \caption{Example of a compatibility graph of a \kepg}\label{fig:KEPmodelc}
\end{figure}

We model a \kepg as a directed graph by creating one vertex for each participant and one arc for each possible transplant.
Formally, the set $\pairset$ contains one vertex for each patient-donor pairs and the set $\altruistset$  one vertex for each altruistic donor.
To construct the \textit{compatibility graph}  \boxm{\compgraph = (\vertexset=\pairset\cup\altruistset,A)}, we add an arc $a = {(uv)}$ between \boxm{u \in V} and \boxm{v \in \pairset} if the kidney of donor $u$ can be transplanted to patient $v$.
A weight function \boxm{\w: A \rightarrow \mathbb{R}^+}  represents the medical benefit of each possible transplant.
Note that determining the weight function is an upstream work and that $\w$ is an input in our case.
This graph is generally quite sparse as it is rare for a patient and a donor to be compatible.
Figure \ref{fig:KEPcgraph} shows an example of compatibility graph and differentiates altruistic donors (orange diamonds) from pairs (red circles).

An {exchange} is a subgraph of $\compgraph$ which represents either a cycle of donation between pairs or a domino chain initiated by altruistic donors.
In the compatibility graph, exchanges are  elementary cycles of length at most $\k$, called \textit{valid cycles} and elementary paths starting by a vertex of $\altruistset$ and having at most $\l$ vertices, \textit{valid paths}. 
A valid cycle could have several symmetrical representations but they are eliminated by restricting the first vertex of the cycle vector to have the lowest identifier.
Thus, a valid cycle $c$ is represented by a unique vector of vertices $(v_1,...,v_{|c|})$ such that \boxm{v_1 < v_j \  \forall j \in \{2,...,|c|\}}.

$\cycleset$ is the set of all valid cycles, $\pathset$ the set of all valid paths and $\exchangeset= \cycleset\cup\pathset$ the set of all possible exchanges.
We refer to the set of vertices (resp. edges) of an exchange $e$ as $V(e)$ (resp. $A(e)$).
The weight of an exchange $e \in \exchangeset$ is $w(e):= \sum\limits_{a \in A(e)} w_a$.
In Figure \ref{fig:KEPcgraph} for example, by taking $\k=3$ and $\l =4$, there exists 8 exchanges: two cycles ($e_1=5-7-6$; $e_2=4-6$) and six paths ($e_3=1-3$; $e_4=1-3-5$; $e_5=1-3-5-7$; $e_6=2-3$; $e_7=2-3-5$; $e_8=2-3-5-7$).

%



Formally, the \kep is a maximum-weight set packing problem, where the considered sets are the exchanges.
As each agent can give or receive at most one kidney, the exchanges must indeed be pairwise disjoint.
Figure \ref{fig:KEPmodelc} shows the optimal solution of the \skep in our example. 
In the \exf (\sexf), each exchange is associated with one binary variable indicating if it is chosen or not in the solution and a unique set of constraints \eqref{exf:phy} is required to model the disjonction of exchanges:

\boxm{\forall e \in \exchangeset, \ x_e = \left\{ \begin{aligned} &1
      \text{ if exchange $e$ is chosen } \\ &0 \text{ otherwise} \end{aligned} \right. }
      \begin{align}
  \quad z^* = \max \sum\limits_{e\in\exchangeset} w_e x_e & \label{exf:obj} \\
  \sum_{\substack{e\in\exchangeset : \\ i\in V(e)}} x_e &\leq 1& \ \forall i \in V \label{exf:phy}  \\ 
   x_e &\in \{0,1\} & \forall e\in \exchangeset \label{exf:varx}
\end{align} 

The number of variables of \sexf grows  exponentially with $\k$ and $\l$, so even computing its linear relaxation $z^*_{LP}$ may need an excessive amount of time or be impossible due to memory issues. 
However, the quality of its linear relaxation makes  \sexf very promising. 
It is actually, up to now, the tightest formulation for the \skep, including compact and other large-scale formulations~\cite{dickerson2016position,mak2017kidney}.
Moreover, these other integer programs are more complex and also suffer from scalability issues.

To overcome the exponential growth of \sexf, its linear relaxation \sexf can be solved by a column generation approach.
It constructs iteratively a (small) set of variables guaranteeing that an optimal solution uses only these variables with the following two steps:
\begin{enumerate}
\item Solve the restricted master problem (RMP): \sexfl restricted on a
  subset $\exchangeset' \subseteq \exchangeset$.
\item Solve the pricing problem: find an ``interesting'' exchange to add in $\exchangeset'$ (go to 1.) or prove none exists (end).
\end{enumerate}

When the restricted master problem is solved, it computes for each vertex $v$ the dual values $\alpha_v$  associated with constraints \eqref{exf:phy}.
The pricing problem of the \exf aims at finding a new exchange with a \textit{positive reduced cost} or proving that none exists.
The reduced cost of an exchange $e$ is given by \boxm{rc_e = w_e - \sum\limits_{v\in V(e)} \alpha_v}.
For a non-basis variable, it estimates the improvement of the objective function if a solution includes $e$, \ie $x_e >0$. 
It is important to note that these reduced costs are in $\mathbb{R}$ and thus can be positive or negative.
As there are two kinds of exchanges, we can decompose this pricing problem into two subproblems:
 \begin{itemize}
    \item The cycle pricing problem: find a cycle of length at most $\k$ of positive reduced cost, or prove none exists.
    \item The path pricing problem: find an elementary path of length at most $\l$ starting by an altruistic donor and with a positive reduced cost, or prove none exists.  
\end{itemize}

The cycle pricing problem can be solved in polynomial time with a Bellman-Ford algorithm.
On the contrary, the path pricing problem is \npc and the proof, based on a reduction from the directed Hamiltonian path problem, was recently given by Plaut \al~\cite{plaut2016hardness}.
We handle this decision problem with algorithms solving the associated optimization problem: the {\empplc} ({\sempplc}).
This article is dedicated to this problem and how to solve it.

\section{Finding elementary paths \label{sec:ELPPRC}}

%
%
%
%
%
%
%
%

The \empplc belongs to the well-known family of paths problems.
It is a special case of the \espprc.
However its specificity---the length constraint---can be exploited to strengthen existing algorithms.

\subsection{Description of the problem}

Let $G=(V,A)$ be a digraph such that the set of vertices $V$ includes a source $s$.  
Each arc $(ij)\in A$ has a cost $c_{ij}$. 
Let $\l$ be the limit on the length (number of arcs) of a path.
An \emphdefw{$(l,i)$-path} \boxm{p=(s,i_1,...,i_l=i)} is an elementary path of length $l$ starting from the source $s$ and ending in $i$.
We denote by $A(p)$ (resp. $V(p)$) the set of arcs (resp. vertices) of $p$ and by \boxm{c_p = \sum\limits_{(ij) \in A(p)} c_{ij}} its cost.
The objective of the \emphdefw{\empplc} (\emphdefabr{\sempplc}{\empplc}) is to find $p^*$ an elementary $(l,i)$-path of minimum cost $c^*$ such that $l\leq \l$.

\subsection{\sempplc in the \skep}


To cast the pricing problem of the \exf as a \textbf{minimization problem}, we consider a new weight function $c$ associating with each arc the \textbf{opposite} of its estimated reduced cost: \boxm{\forall (uv) \in A, c_{uv} =-w_{uv} + \alpha_v }.
We also construct a new directed graph containing an artificial source $s$ linked to each altruistic donor \boxm{\compgraph' = (V\cup\{s\},A')} where \boxm{A' = A \cup \{(su) \ \forall u \in N\}}.
The function $c$ is extended to these new arcs:  \boxm{\forall u \in N, c_{sv} = \alpha_v}.
A valid path contains $\l+1$ vertices (including the source) and $\l$ arcs in $\compgraph'$.
For an exchange $e$, \boxm{c_e = \sum\limits_{(uv)\in A'(e)} c_{uv} =  \sum\limits_{v \in V(e)} \alpha_v - w_e = -rc_e}.
As \boxm{c_e = -rc_e}, the pricing problem aims at finding paths of \textbf{negative weight} or prove none exists.
Thus, solving  \sempplc on $\compgraph'$ provides an answer for both cases.

\subsection{Standard approaches}

Interest for elementary shortest path problems mainly arose from vehicle routing applications solved by \cg.  
The elementarity constraint is often relaxed to get a simpler problem, which can be relevant in many applications such as vehicle  routing problems where a vehicle can visit the same place twice.
In the standard case, one just wants to solve the \defw{shortest path problem} (\defabr{SPP}{shortest path problem}), which can be done with a Bellman-Ford algorithm.  
In the presence of resource constraints, such as time windows or vehicle capacities, the problem becomes harder.
In 1988, Desrochers~\cite{desrochers1988spprc} proposed an extension of the Bellman-Ford algorithm for the \defw{shortest path problem with resource constraints}
(\defabr{SPPRC}{shortest path problem with resource constraints}).
However, Feillet \al~\cite{feillet2004espprc} argued that relaxing the elementarity constraint can lead to bounds of poor quality, thus proposed to extend Desrochers' algorithm in order to solve the \nphard  \emphdefw{\espprc} (\emphdefabr{\sespprc}{\espprc}).  
In the \sespprc,  paths have limited resources  (instead of having a maximum length as in \sempplc).
Let $R$ be the number of resource types and \boxm{g_{ij}^r \geq 0} the consumption of resource $r$ along the arc $(ij)$.
Each vertex $i\in V$ constrains the path to reach it with a resource consumption belonging to \boxm{\left[a_i^r,b_i^r\right]} for each resource $r$.
The objective is to find a path of minimum cost $c$ such that every resource constraint is satisfied.
By considering a single resource with a unit consumption and by setting the bound on this resource consumption to the length limit, \sespprc describes \sempplc. 
Formally let $R = 1$ and, \boxm{\forall i \in V: } \boxm{a_i = 0} and \boxm{b_i = \l}.
In addition we set \boxm{g_{ij} = 1}, \boxm{\forall (ij) \in A} and observe that \sempplc is a special case of \sespprc.

When dealing with the \sempplc, like for any problem, the objective is generally to find the optimal solution.
Exact algorithms are designed for this, but as \sempplc is \nphard, they might take a long time to get it.
On the other hand, in a \cg framework it is not important to find optimal solutions, but to find a solution with the same sign as the optimal solution.
Indeed, assume the pricing problem is to find a path with a \textbf{negative cost}.
Then any feasible solution of \sempplc is a new column to add to the restricted master problem.
Similarly, if a lower bound has a positive (or zero)  cost, the optimality proof of the linear relaxation is done.
Lower bounds and feasible solutions can be computed with relaxations and heuristics respectively.

Numerous approaches solving paths problems, including Desrochers' and
Feillet', are based on dynamic programing and labeling algorithms
following Held and Karp results~\cite{held1962dynamic}.
This dynamic program solving the \sempplc has a space complexity of $\mathcal{O}(|V|2^{|V|})$, a time complexity of $\mathcal{O}(|A|2^{|V|})$ and does not scale up to large instances. 
A common idea to overcome scaling issues is to relax the problem in order to deal with a smaller search space.  
Relaxing a minimization problem aims at quickly providing lower bounds. 
If we relax the constraint of elementarity, the search space is strongly reduced since the visited vertices are not remembered anymore.
We can also relax the resource constraints to solve the shorstest path problem.
More complex dynamic programs relax only partially these constraints.
This is the case of the \ngr relaxation proposed by Baldacci \al~\cite{baldacci2011ngroute} which constructs partially elementary paths.
We present their algorithm and how we reinforce it for our problem in Section~\ref{sec:empplc-NG}.
Note that in the kidney exchange context, the elementarity constraint cannot be relaxed to get a feasible solution, but it can be relaxed in the RMP, ``temporarily'',  in order to speed up the \cg.
Actually, compatibility graphs are generally sparse,  unlike graphs of vehicle routing problems which are usually complete, and relaxed solutions may be elementary anyway. 

Another, and opposite, idea to reduce the search space in dynamic programming is to restrict the problem.
More constrained problems will provide feasible solutions and thus upper bounds on the \sempplc optimal solution.
We introduce the \cd algorithm proposed by Alon \al~\cite{alon1995color} and present our contributions in Section \ref{sec:empplc-CD}.

\bigskip

\correction{

These two dynamic programs, the \cd and the \ngr, are the core of our study of the \sempplc. 
\corbis{They provide a different  type of solution and to assess their interest, we compare them to two alternative algorithms that serve as baseline methods.}


}

Firstly, we adapt an integer linear programming formulation of the \tsp called the \emphdefw{\tsf} (\emphdefabr{\stsf}{\tsf}), which was proposed by Fox, Gavish and Graves~\cite{fox1980n} for the \tsp. 
In this model, a decision variable $y^l_{ij}$ states if arc $(ij)$ is taken at the $l^{\text{th}}$ position of the path.
The integer program \stsf is used to compute the optimal solutions of the different \sempplc instances.
It is also solved within a time limit of 1 second, giving the best lower and upper bounds \emphdefabr{\iplb}{best lower bound of \stsf in 1 second} and  \emphdefabr{\ipub}{best upper bound of \stsf in 1 second}

\begin{figure}

  \centering
  \begin{subfigure}{0.45\textwidth}
    \includegraphics[scale=0.55,page=1]{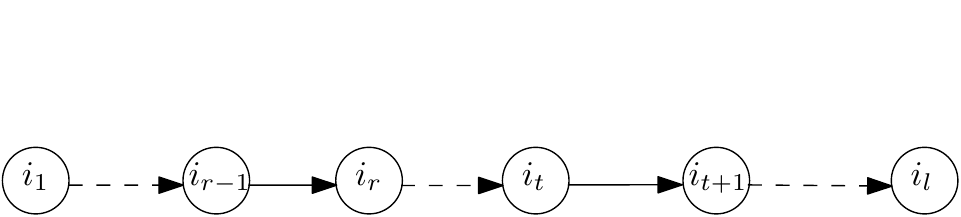}
    \caption{Initial path}
  \end{subfigure}
  \begin{subfigure}{0.45\textwidth}
    \includegraphics[scale=0.55,page=4]{EMPPLC-LS.pdf}
    \caption{Insertion}
  \end{subfigure}

 \begin{subfigure}{0.45\textwidth}
   \includegraphics[scale=0.55,page=2]{EMPPLC-LS.pdf}
   \caption{Suppression}
 \end{subfigure}
 \begin{subfigure}{0.45\textwidth}
   \includegraphics[scale=0.55,page=3]{EMPPLC-LS.pdf}
   \caption{Exchange}
 \end{subfigure}
 \captionsetup{justification=centering}
 \caption{Movements of the local search\label{fig:ls}.\newline Dashed arrows represent subpaths, plain arrows are edges and crossed arrows are removed in the movement.}
\end{figure}

Secondly, we develop a quick heuristic based on local search: the algorithm moves from the current solution $p$ to a better solution $p'$ in its neighborhood.
This solution $p'$ is constructed by inserting, removing or exchanging up to 3 vertices in $p$ (see figure \ref{fig:ls}) and kept if $c_{p'} < c_p$.
The solution with the minimum cost is returned after one second (solution \emphdefabr{\lm}{best solution of local search in 1 second}), as well as the first path of negative cost (solution \emphdefabr{\lf}{first negative solution of local search}).
If no path of negative cost is found in the time limit, then \lf is equal to \lm.


\subsection{Exploit the length constraint \label{sec:empplc-dist}}


As we look for a path of length at most $\l$, we can use this information to reinforce shortest path algorithms.
We compute with Floyd-Warshall algorithm the distance function
${d : V\times V \rightarrow \mathbb{N}\cup\{+\infty \}}$ where $d(i,j)$
is the shortest path between $i$ and $j$, with respect to the
\textbf{number} of arcs.
We define the \emphdefw{extended neighborhood} \defs{\gamma(i)}{extended neighborhood} of vertex $i$
as the set of vertices that may appear in any path starting at the
source of length at most $\l$ including $i$:
\boxm{\gamma(i) := \{ j\in V: d(s,i)+d(i,j)\leq \l \text{ or } d(s,j) +
d(j,i) \leq \l\}}.  
Note that if ${i \in \gamma(j)}$ then ${j\in \gamma(i)}$.
We define the \emphdefw{extended predecessors}  \defs{\gamma^-(i)}{extended predecessors} of vertex
$i$ as the sets of vertices that can reach $i$ in a path of length at
most $\l$: \boxm{\gamma^-(i) := \{ j\in V: d(s,j) + d(j,i) \leq \l\}}.

\corbis{The} distance function is used first to perform a \textbf{preprocessing} on the graph $G$ by removing every arc (and vertex) that is too far from the source to be contained in a path of length $\l$.
The sets of extended neighbors and predecessors also take part in the improvements of the next sections.


%
%


\section{The color coding restriction \label{sec:empplc-CD}}


The randomized algorithm referred to as the \cd was introduced by Alon \al~\cite{alon1995color} to solve the subgraph isomorphism problem.
It can be applied in particular to solve the \empplc.
Its principle is to randomly assign a color to each vertex, then to solve a dynamic program finding the path of minimum weight whose vertices all have different colors.
Such a path is said to be \textbf{colorful} and searching for it is computationally more efficient than the search for an elementary path as there are $C$ distinct vertex identifiers instead of $|V|$.
However, this colorful path is not optimal in general.
Indeed, for an optimal solution to be found with \cd, the random coloring must, by chance, assign different colors to its vertices.
Thus, the two steps of the algorithm are repeated so that each solving of the dynamic program might return a different path, increasing the probability that one of them is optimal.

In all its applications and theoretical studies, the coloring step is performed with a discrete uniform distribution. 
Only Kneis \al~\cite{kneis2009derandomizing} proposed another probability distribution to color the graph, but for a derandomization purpose.
We introduce in this section new strategies for this step of the algorithm and focus on how this strategies are helpful to efficiently solve the \sempplc for our pricing step.
We refer the reader to Pansart \al~\cite{pansart2020ecai} for the theoretical study of these new strategies in the general context of subgraph isomorphism problem, but also for an extensive literature review on \cd and proofs of properties mentioned in the following.


%
%
%
%

\subsection{Description of the algorithm}

To solve the \empplc in a weighted graph $G=(V,A)$, the \cd uses two steps:
\begin{enumerate}
\item Coloring: randomly assign a color $c \in \{1,...,C\}$ to each vertex $v\in V\setminus\{s\}$
\item Dynamic programming: finds the colorful path of minimum weight, of length at most $\l$, whose vertices all have distinct colors. $f^*(\mathcal{C},i)$ is
  the minimal cost of a path from $s$ to $i$, using $|\mathcal{C}|-1$
  arcs and visiting vertices of each color of $\mathcal{C}$:
\begin{equation}
    f^*(\mathcal{C},i) = \min\limits_{\substack{j\in N^-(i)\\ c_j \in \mathcal{C}\setminus \{c_i\}}} \{ f^*( \mathcal{C}\setminus \{c_i\},j) + c_{ji} \}
 \end{equation} 
\end{enumerate}

%

These two steps are repeated several times in order to guarantee a probability high enough that one of these trials finds an optimal path.
In the standard version of the \cd, the coloring step applies a discrete uniform distribution to assign the colors and $C = \l$.
In this case, the number of trials required to ensure that a specific path (\eg an optimal path) of length $\l$ is colorful with probability $\rho$ is \boxm{t = \frac{ln(1-\rho)}{ln(1-\frac{C!}{C^C})}}.

Using more colors can however increase the chance for a path to be colorful, thus reduce the number of trials $t$, but it also increases the complexity of the dynamic programming step.
H\"uffner \textit{et al.}~\cite{huffner2008algorithm} analyzed the best trade-off between these two parameters: the number of trials and the runtime of one trial.
Another lever on the \cd efficiency is the probability distribution used to color the graph. 
Surprisingly though, the coloring strategy always follows the discrete uniform distribution, except in Kneis \al~\cite{kneis2009derandomizing} for a derandomization purpose.
With this strategy, assigning a color to a vertex does not depend on other vertices colors.
Yet, using a different probability distribution can significantly increase the probability that an optimal path is colorful, and, contrarily to increasing $C$, will not affect the computational performance of the dynamic program.
The new method we introduce in the next section is based on this idea.

\subsection{New randomized strategies}

We propose to use a coloring technique that aims at spreading the colors in extended neighborhoods.
To do so,  it tries to make extended neighborhoods colorful by relying on three main ideas.
First, a preprocessing step applies a local search to create an ordering of vertices $x_1,...x_n$ ($x_i \in V$) in which extended neighbors are gathered.
Secondly, the vertices are colored by intervals of size  $C$ such that each interval is colorful.
The color assigned to a vertex therefore depends on the color of other vertices in this interval.
Finally, the ordering is shifted so that the intervals are  made up of different extended neighbors each iteration.


The preprocessing step constructs a \textbf{coloring sequence} with the objective to gather in this sequence vertices that might belong to a solution.
Such vertices are identified thanks to the extended neighborhoods.
Let $x_i$ be the position of vertex $i$ in the sequence and $\delta$ be the sum of differences between the positions of two extended neighbors: \boxm{\delta  = \sum\limits_{\substack{ i \in V\\ j \in \gamma(i) }}|x_i - x_j|}.
Our preprocessing constructs a coloring sequence such that $\delta$ is minimized, \ie the average distance between the positions of two extended neighbors is minimized.
This problem is known as the (minimum) linear arrangement problem~\cite{adolphson1973optimal} that was proven to be \npc~\cite{garey1974some}.
We handle it with a local search and this method is referred to as \lssum.

%
%
%

We denote by $(x_1,...,x_n)$ the coloring sequence of  vertices of $V\setminus\{s\}$ provided by this preprocessing step.
Our coloring strategy \pmax splits the coloring sequence into intervals of size $C$ and colors vertices such that these intervals are colorful.
Concretely, the colors vector of vertices in an interval $I$ is a permutation of $\{1,...,C\}$ drawn with  a uniform distribution over all the permutations.
The color of a vertex thus depends on the colors taken by other vertices of the same interval, but on no other vertex.
Thus, two vertices in the same interval cannot have the same color and two vertices in different intervals are assigned the same color with probability $\frac{1}{C}$ (proof in~\cite{pansart2020ecai}).
Once the dynamic program is solved, the coloring sequence is offset by 1 to the left.
So after the first trial, the coloring sequence is $(x_2,...,x_n,x_1)$.
This \pshift strategy means that, after $C$ trials of the \cd, every subsequence of size $C$ has been colorful once.

Thus, if an optimal path has been gathered enough in the coloring sequence by the preprocessing, it will be colorful and found in $C$ trials only.
In general we cannot know when this happens, but it can be guaranteed by a parameter of the coloring sequence: the \textbf{maximum} distance  between the positions of two extended neighbors  \boxm{\Delta  = \max\limits_{\substack{ i \in V\\ j \in \gamma(i) }}|x_i - x_j|}.
When $\Delta \leq C$, every optimal path is contained in a subsequence of size at most $C$ and our algorithm guarantees to find them with a probability of 1 with at most $C$ trials of the \cd, so only $C$ calls to the costly dynamic program.

The worst case for this coloring strategy is to have the positions of all the vertices of every optimal path in different intervals. 
In this case, which should not happen often in practice thanks to the \lssum, the probability that a path is colorful is the same than when using the discrete uniform distribution (proof in~\cite{pansart2020ecai}).
Thus, the method we propose always improve the chance to find an optimal solution over the original \cd.

To sum up, our method \oands (\lssum + \pshift) contains a preprocessing phase and a \cd phase with three steps:
\begin{enumerate}
\item Apply a \pmax coloring strategy
\item Solve the dynamic program
\item Shift the coloring sequence
\end{enumerate}

When $\Delta \leq C$ these three steps are repeated only $C$ times and the solution is optimal, otherwise we limit the running time and get the best solution after one second (solution \textbf{\cm}).
We  also return the first negative solution (solution \textbf{\cf}), if applicable.
This time limit is small as the \empplc must be solved \corbis{a lot} of times in a \cg scheme.
On the contrary, the \lssum is a preprocessing applied only once by instance of the master problem, so we let the local search runs for 300 seconds.
When applied in the pricing step to solve the \kep, this method often returns the optimal solution, even when $\Delta > C$.
Indeed, graphs are sparse so the extended neighborhoods are small enough to make the preprocessing powerful.
This performance allows a quick and effective solving of the pricing problem.
Yet, if the \cd does not find a path of negative cost, it does not prove that the \cg has finished since this solution is only an upper bound.
Thus, another method is still required to make this proof: this is the role of the \ngr relaxation.

\pagebreak
\section{The ng-route relaxation \label{sec:empplc-NG}}
\corbis{In 2011}, Baldacci \al~\cite{baldacci2011ngroute} proposed a new relaxation of the \espprc called \emphdefw{\ngr}. 
The memory of a path is relaxed so that the search space of the dynamic program is reduced.
In practice, a path constructed in the \ngr relaxation, called an \ngp, can forget that it went through some vertices and may visit them several times and be non elementary.
However, if it turns out that the \ngp is elementary, then it is an optimal solution of the \sespprc.
We describe the adaptation of this algorithm for the \sempplc special case.

Note that solving the \sempplc with the \ngr relaxation in  a \cg scheme can lead to the introduction of non elementary columns in the RMP.
In this case, the \cg does not solve the linear relaxation of the master problem, but a relaxation of this linear problem. 
In the \skep context, the solution obtained with non-elementary paths is thus an upper bound on $\zexfl$, but it can be used similarly, for example to assess the quality of a feasible solution.
Of course if only elementary paths were added by the \ngr relaxation, this upper bound is actually $\zexfl$.
As compatibility graphs are rather sparse, the elementarity will be often satisfied by \ngps.

\subsection{Description of the algorithm}
In this relaxation, each vertex $i$ has a ``memory'', also named \emphdefw{\ngs}, denoted by \boxm{\eta_i \subseteq V} and such that \boxm{i \in \eta_i}.
If an \ngp goes through $i$, it can remember only vertices of $\eta_i$.  
As it is true for each vertex, an \ngp only remembers the vertices appearing in every \ngs.
An \ngp of minimum cost with respect to these \ngss can be constructed by a dynamic program, either in a forward or backward scheme.
We detail below the forward dynamic program and  refer the reader to the paper of Baldacci \al for the backward version.

Each path
\boxm{p=(s,i_1,...,i_l)} is associated with a set of remembered vertices 
\boxm{\Pi(p)= \left\{ i_r : i_r \in \bigcap\limits_{t=r+1}^{l}\eta_{i_t},
  r=1,...,l-1 \right\} \bigcup \{i_l\}}. 
  
A forward \emphdefw{\ngp} \boxm{(\Pi, l, i)}, is a (non necessarily elementary) path
\boxm{p=(s,i_1,...,i_l = i)} starting from $s$, ending in $i$, using $l$ arcs and such that
\boxm{\Pi = \Pi(p)}. 
$\Pi$ represents the memory of $p$, since no vertex of $\Pi$ can be used to extend $p$. 
$p$ is constructed by adding $i$ to a smaller \ngp that belongs to the set \boxm{\Psi^- \left(\Pi, l, i \right)}:

\begin{align*}
  \Psi^- \left(\Pi, l, i \right) = \left\{  \right.
  &(\Pi', l-1, j) \text{ ng-paths s.t.}: \\ 
  \ &j\in N^-(i),
   \Pi = \left(\Pi' \cap \eta_i \right) \cup \{i\},
    \Pi'\subseteq \eta_j,
      j\in \Pi',
       i \notin \Pi'
      \left.  \right\}
\end{align*}

$f^*(\Pi, l, i)$ is the minimal cost of an \ngp $(\Pi, l, i)$ and
can be computed with the following recursive formula:
\begin{equation}
  f^*(\Pi, l, i) = \min\limits_{(\Pi', l-1, j) \in \Psi^- \left(\Pi, l, i \right) } \{ f^* \left(\Pi', l-1, j \right)+ c_{ji} \}
\end{equation}

\subsection{Decremental State-Space Relaxation}

Pecin \al~\cite{pecin2013ngroute} proposed to use the Decremental State-Space Relaxation (DSSR) technique of Righini and
Salani’s~\cite{righini2008new}, an iterative algorithm in which the search space is even more relaxed than in the pure \ngr relaxation. 
At each iteration $k$, each vertex $i$ is associated with a set $\mu^k_i$ that takes the role of the \ngs $\eta_i$ in the dynamic program.
At the first iteration the subsets $\mu^0_i$ are empty sets.
When the solution $p_k$ of iteration $k$ is not elementary and does not respect the chosen criterion, vertices are added to the sets  $\mu^k_i$ and a new iteration begins.
There are three main possible criteria in the DSSR \ngr algorithm.
\begin{itemize}
\item {predefined}: original \ngss $\eta_i$ are computed and the DSSR continues until $p_k$ is either elementary or a feasible \ngp with respect to these sets.
Vertices are added to sets  $\mu^k_i$ only if they belong to $\eta_i$.
\item {limited}: the DSSR continues until $p_k$ is elementary or the sizes of sets $\mu^k_i$ exceed a given limit.
\item {unlimited}: the DSSR continues until $p_k$ is elementary.
\end{itemize}

Without descent or with a {predefined} one, \ngss are the heart of the algorithm since they determine the quality of the solution as well as the computation efficiency.
  When $\eta_i$ is empty for every vertex, there is absolutely no constraint on the elementarity of the path and the \ngr relaxation solves the SPPRC.  
When $\eta_i = V$ for every vertex, the \ngr is not a relaxation anymore and the solution is necessarily elementary.
Except for the unlimited DSSR-\ngr, the \ngss have a limited size of $\Lambda$ and, in general, they are constructed randomly.
We propose to exploit the length constraint of the problem and to take into account the extended neighborhoods in the  \ngss construction. 
Indeed, it is sufficient for a vertex to ``remember'' in $\eta_i$ only its extended predecessors since they are the only vertices that can appear in a path reaching $i$. 
Moreoever, when the \sempplc is embedded in a \cg framework, vertices are associated with a dual value which usually represents the interest for the vertex to appear in the solution. 
The memory of a vertex can therefore be composed by its $\Lambda$  extended neighbors with the highest dual value.

\correction{
We also apply the filtering proposed by Pecin \al~\cite{pecin2013ngroute} based on the alternation between forward and backward computations of the \ngp in a DSSR scheme. 
Assume we have $UB_p$ an upper bound on the \sempplc optimal value.
At each iteration, the costs computed in the previous iteration can be used as completion bounds.
Thus, for each state of the dynamic program, a lower bound on the cost of a \ngp constructed from this state can be calculated.
If this bound is greater than $UB_p$, then the state is pruned as it never be part of an optimal solution.
A natural upper bound in our case is $0$ as we are looking for a path of a negative cost.

The solution of this relaxation can be a non elementary path.
However in the context of the \kep, graph are sparse so the elementarity is often achieved even with a relaxed memory.
Moreover, the \ngr algorithm is not aimed at finding columns to add in the \cg, but rather at proving the end of the \cg.
Ideally, it will be called only once at the end of the \cg and in practice the number of calls is indeed low (4.2 in our experiments) due to the efficiency of our \cd method.
}

\section{Experiments on \sempplc algorithms\label{sec:empplc-exp}}

We first detail the instances and the protocol, then we analyze the performance of the different algorithms studied in this article to solve the \empplc.

%
%
%
%
%
\subsection{Instances and protocol \label{sec:empplc-protocol}}
%
%
Experiments are conducted on several \sempplc instances generated from the pricing step of \skep instances.
Pools of patients and donors are created using an online\footnote{available at \url{http://www.dcs.gla.ac.uk/~jamest/kidney-webapp/\#/generator}}  Saidman-based generator~\cite{saidman2006increasing} with realistic parameters, leading to sparse graphs.
In this benchmark called  \kbear, there are  27 different classes of instances.
In particular, the number of patients varies between 50 and 250, \boxm{\k =3} and \boxm{\l \in \llist}.
The \exf is solved by \cg on one instance of each class and \sempplc instances are extracted from the first, last and middle iterations of the pricing problem. 
Note that the 54 instances generated from the first and middle iterations (instances  \ekbearm) contain a solution of negative weight while the optimal value for the 27 last iterations (instances \ekbearz)  is zero\footnote{all the 81 instances are available at \url{https://pagesperso.g-scop.grenoble-inp.fr/\~pansartl/data/instances-EKBR.zip}}.
The purpose of generating instances with such a procedure is to get dual values at different stages of the \cg, so different weight functions $c$.
%
%

In a \cg, methods providing feasible solutions are designed to quickly find new columns to add, \ie find a path with a negative reduced cost.
For this reason, heuristics either stop when the first solution of negative cost is found, or run within a small time limit.
On the contrary, relaxations must find good solutions to allow filtering and computation of good bounds.
Therefore, the \ngr relaxation is not limited in time while the \cd and local search algorithms are set to return the first solution of negative cost and the best solution after 1 second of running time. 
Note that the \cd actually ends after at least one trial was completely executed, so the effective running time may exceed this time limit.
The integer program is used to compute the optimal solution, but also best upper and lower bounds within 1 second.
All in all, seven solution types are reported:
\begin{tight_itemize}
\item 5 upper bounds
\begin{tight_itemize}
\item \cf: first solution of color coding
\item \cm: best solution of color coding in 1 second
\item \lf: first solution of local search 
\item  \lm: best solution of local search in 1 second
\item \ipub: best feasible solution of the integer program in 1 second
\end{tight_itemize}
\item 2 lower bounds
\begin{tight_itemize}
\item \ngm: best \ngr of the instance
\item  \iplb: best lower bound of the integer program in 1 second
\end{tight_itemize}
\end{tight_itemize}


The quality of a solution of cost $c$ is evaluated against the optimal value $c^*$ with three performance indicators:  its gap to optimal (\boxm{\frac{|c - c^*| }{c^*}}), its optimality (\boxm{c = c^*}) and its sign compared to optimal solution.
Since instances of \ekbearz have an optimal value of 0, solutions \cf and \lf are not reported for these instances and the gap is not computed either.

\corbis{All the experiments were performed on an Intel Xeon E5-2440 v2 @ 1.9 GHz processor and 32 GB of RAM.}

%
%

\subsection{\Cd}

{\captionsetup[subfigure]{justification=raggedright,singlelinecheck=off}

\begin{figure}[ht!]
\begin{minipage}{0.74\columnwidth}

\subcaption{$\l = 4$ (18 instances)}
\includegraphics[scale=0.8]{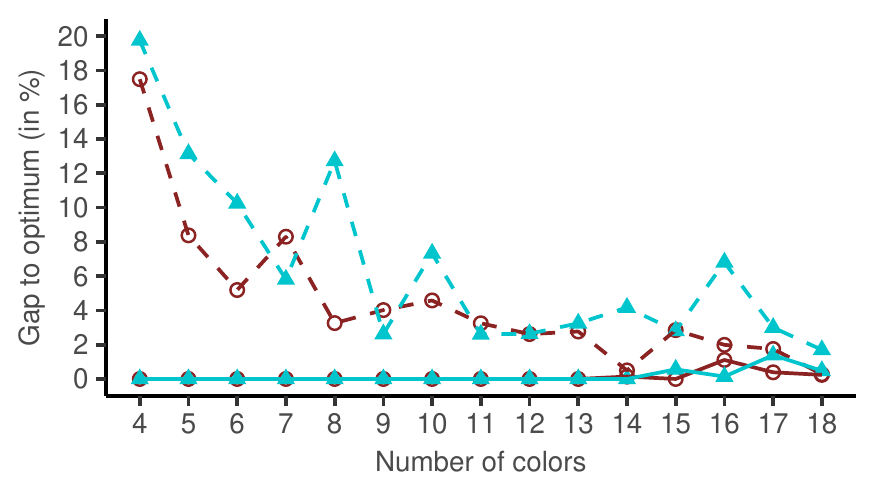}

\subcaption{$\l = 7$ (18 instances)}
\includegraphics[scale=0.8]{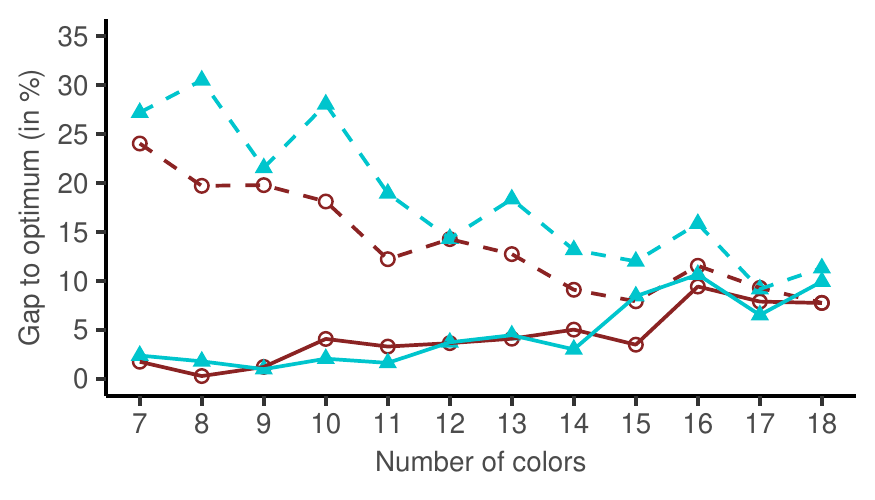}

\subcaption{$\l = 13$ (18 instances)}
\includegraphics[scale=0.8]{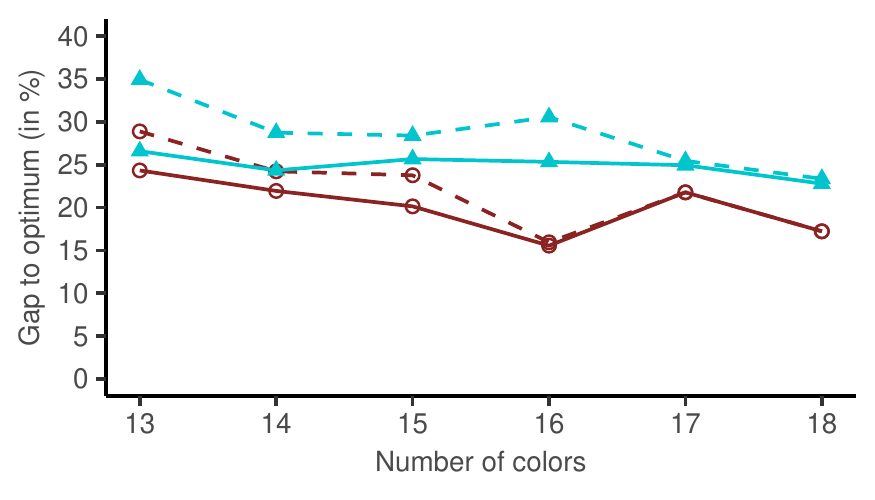}

\end{minipage}
\begin{minipage}{0.25\columnwidth}
\includegraphics[scale=1]{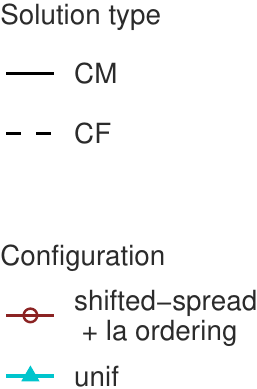}
\end{minipage}
 \captionsetup{justification=centering}
\caption{Gap between CF or CM and OPT depending on the configuration and the number of colors, for the \ekbearm benchmark\label{fig:cd-gap}}
\end{figure}
}

We conducted experiments to determine the best number of colors and the  best probability distribution to solve the \sempplc with \cd.
We observe that our method \oands outperforms the standard \cd algorithm which uses a \prandom distribution.

%

\begin{figure}[htb]
\centering
 \captionsetup{justification=centering}
\includegraphics[scale=0.8]{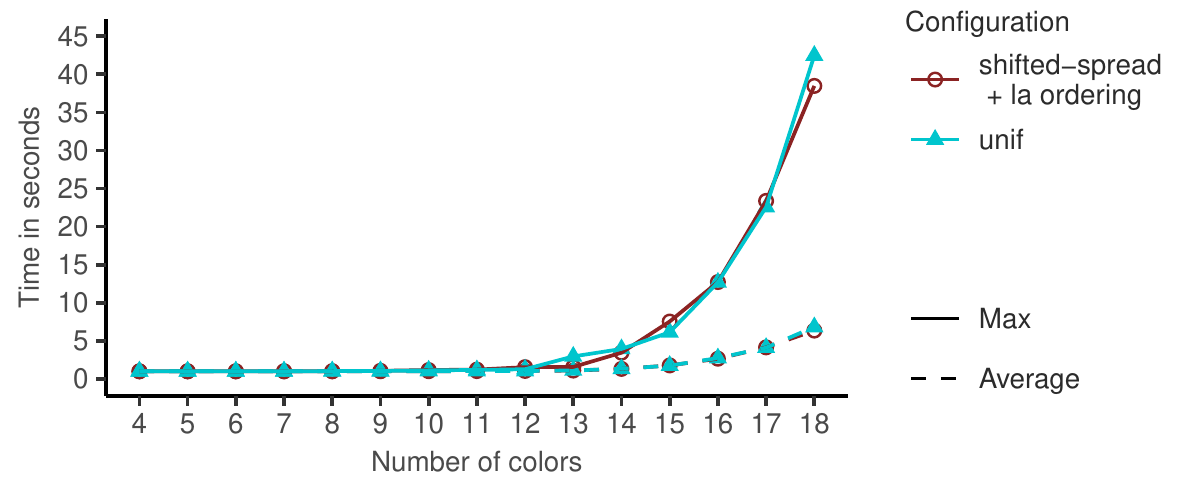}
\caption{Average and maximum \cd running times depending on the number of colors and the configuration on \ekbearm instances\label{tab:empplc-cd-time}}
\end{figure}

Figures \ref{fig:cd-gap} show the gap for the \ekbearm benchmark. 
The 54 instances are grouped in three sets of 18 instances, according to the parameter $\l$. 
The first negative solution \cf for the \oands  strategy is almost always better than for  the standard \cd \prandom.
Its quality increases with the number of colors and this is due to the fact that a trial of \cd is more efficient with more colors, both for \oands and \prandom configurations.
However, an important condition for the  \cd to return good solutions is to make many trials.
As increasing $C$ also increases the running time  of one trial, it leads to poorer solutions in the same time limit, hence the growing gap for \cm.
When $C$ is too big, the \cd can exceed the time limit  (see Figure~\ref{tab:empplc-cd-time}) because an iteration runs during more than one second.
In this case, the \cd actually stops after a single iteration and \mbox{\cf $=$ \cm}.
From this analysis, the best compromise seems to run the \cd with a \oands configuration  and a small number of colors (we take \boxm{C=\l+1}).

\subsection{\ngr}

\begin{table}[htb]
\begin{center}
\begin{tabular}{lcc}
\toprule
&\ngs creation &  Descent  \\
\midrule
\ngub &  {\bfseries\color{MaCouleur}U}niform & {\bfseries\color{MaCouleur}N}one \\
\ngdb &{\bfseries\color{MaCouleur}N}eighborhood & {\bfseries\color{MaCouleur}N}one  \\
\ngl & - & {{\bfseries\color{MaCouleur}L}imited}\\
\ngup & {\bfseries\color{MaCouleur}U}niform& {{\bfseries\color{MaCouleur}P}redefined}\\
\ngdp & {\bfseries\color{MaCouleur}N}eighborhood& {{\bfseries\color{MaCouleur}P}redefined}\\
\bottomrule 
\end{tabular}
\caption{The five versions of the \ngr relaxation\label{fig:empplc-ngconfig}}
\end{center}
\end{table}

Five versions of the \ngr are implemented and tested, voluntarily  omitting the {unlimited} DSSR as it provides no control on the computation time and memory space.
Table \ref{fig:empplc-ngconfig} sums up the different versions and how they construct the \ngs and which DSSR is applied, if any.
We tested different size limits for the \ngs (5 to 13), but it appears that they make no difference on the solution quality.
On the other hand, increasing this size limit deteriorates the computation time, in particular for configurations without descent.
Thus, only results for a size of 5 are kept. 

\begin{figure}[htb]
    \centering
      \begin{subfigure}[b]{1\textwidth}
        \centering \includegraphics[scale=1]{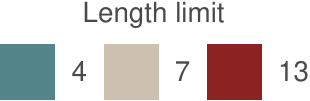}
    \end{subfigure}
    
    \medskip
    
    \begin{subfigure}[b]{1\textwidth} 
        \centering \includegraphics[scale=0.7]{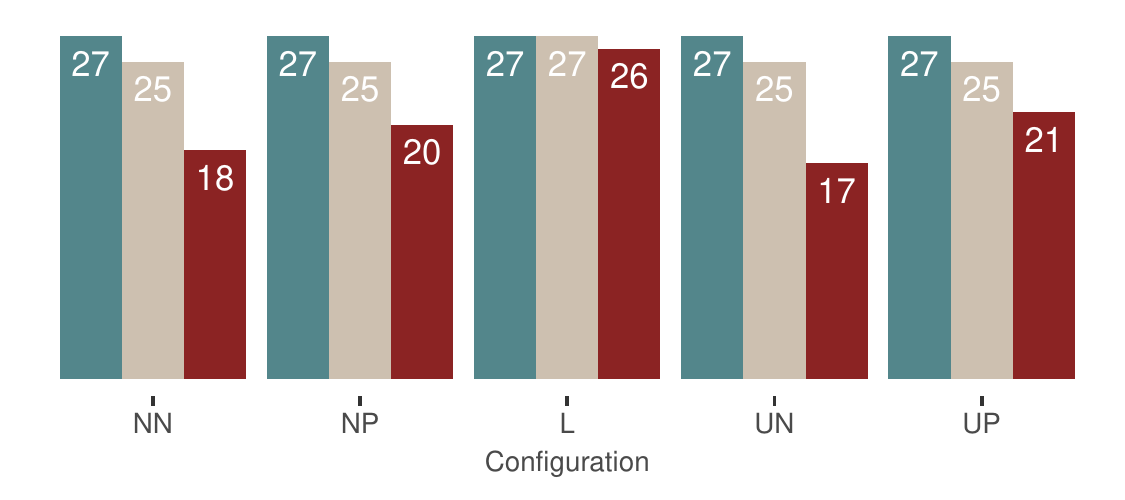}
        \caption{Number of instances, out of 27, for which NG $=$ OPT\label{fig:ng-opt}}
    \end{subfigure}
    
    \begin{subfigure}[b]{1\textwidth}
        \centering \includegraphics[scale=0.7]{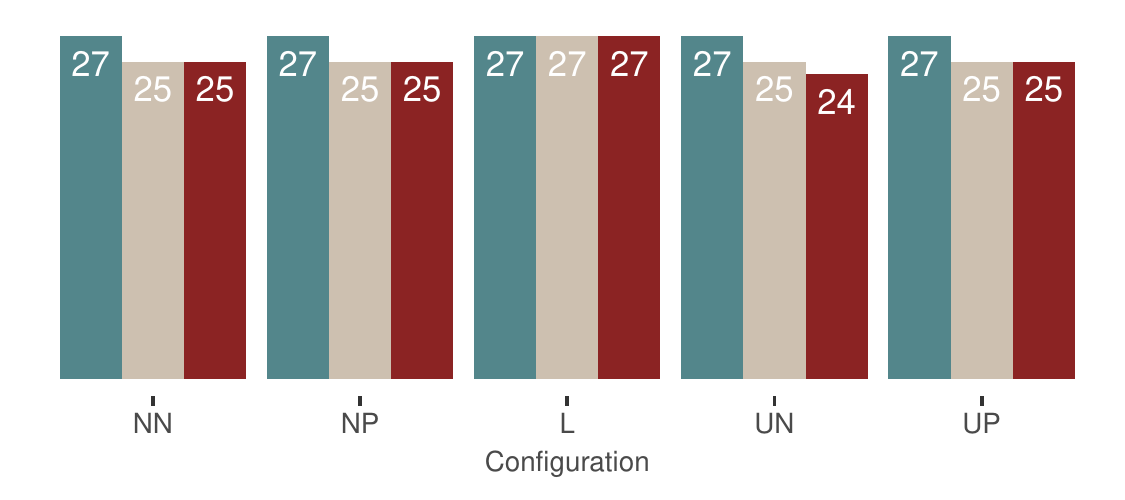}
        \caption{Number of instances, out of 27, for which NG and OPT have the same sign\label{fig:ng-sign}}
    \end{subfigure}   
     \captionsetup{justification=centering}
  \caption{Quality of the \ngr solution depending on the configuration on \ekbear instances
  }\label{fig:ng-res}
\end{figure}

Figures \ref{fig:ng-res} illustrate the quality of the \ngr solution for the \ekbear benchmark. 
The 81 instances are grouped into three sets of 27 instances, according to the parameter $\l$.
Figure \ref{fig:ng-opt} shows the number of instances optimally solved by the \ngr relaxation, \ie the number of instances for which the \ngr returns an elementary \ngp, so the optimal solution.
Figure \ref{fig:ng-sign} shows the number of instances for which the lower bound returned by the relaxation has the same sign as the optimal solution, a success criterion for the \cg algorithm.
Both figures demonstrate the good performance of the {limited} DSSR compared to other configurations, even though all of them are quite efficient.
Still, the {limited} DSSR is the only configuration that always returns a solution of the same sign as the optimal one and finds this optimal solution almost every time.
The fact that the solution of the \ngr is often the optimal solution explains the fact that increasing the \ngs size is not interesting, as even when they are small the solution is elementary.

%
\subsection{Comparisons of four \sempplc algorithms}

The four algorithms were implemented and tested on the two benchmarks \mbox{\ekbearm} and \ekbearz. 
Recall that for \ekbearz, the gap cannot be computed and that \cf and \lf solutions do not make sense as no feasible solutions of negative weight can be found for these instances.
Thus, for this benchmark, Table \ref{tab:empplc-nb} shows only the number of instances for which:
\begin{tight_itemize}
\item the solution has the same sign as the optimal one 
\item the solution is the optimal one
\end{tight_itemize}
for solutions \cm, \lm, \ipub, \ngm and \iplb.

\begin{table}[htbp]
\centering
\renewcommand{\arraystretch}{1.5}
\begin{tabular}{V{3cm}ccccc}
\toprule
& \multicolumn{3}{c}{Upper bounds} & \multicolumn{2}{c}{Lower bounds} \\
\cmidrule(lr){2-4}\cmidrule(lr){5-6} 
 & \cm  & \lm & \ipub &  \ngm & \iplb \\
\midrule 
\# instances with good sign & \bfseries 27 & \bfseries  27 &  24 & \bfseries  27 & 19 \\
\# instances with optimal solution & \bfseries 27 &   26 &   24 & \bfseries  27 & 19 \\
\bottomrule
\end{tabular}
\caption{Quality of solutions for the 27 \ekbearz instances \label{tab:empplc-nb}}
\end{table}

\begin{table}[htbp]
\centering
\renewcommand{\arraystretch}{1.5}
\begin{tabular}{V{3cm}ccccccc}
\toprule
& \multicolumn{5}{c}{Upper bounds} & \multicolumn{2}{c}{Lower bounds} \\
\cmidrule(lr){2-6}\cmidrule(lr){7-8} 
& \cf & \cm & \lf & \lm & \ipub &  \ngm & \iplb \\
\midrule
\# instances with good sign & \bfseries 54 & \bfseries  54 & 41 & 41 &51 & \bfseries  54 &40 \\
Average gap  on such instances (\%) &  17.4 & \bfseries  7.4 & 75.8 & 25.2 &15.2  & \bfseries  0.01 &\bfseries 0 \\
\# instances with optimal solution & 18 &\bfseries 40 & 1 & 17 & \bfseries 40 & \bfseries 53 &40 \\
\bottomrule
\end{tabular}
\caption{Quality of solutions for the 54 \ekbearm instances \label{tab:empplc-gap}}
\end{table}

Table \ref{tab:empplc-gap} for \ekbeam is more complete as it also displays the gap with the optimal solution for instances having a solution of the same sign as the optimal one.
This gap is therefore not computed on the same number of instances for all the solutions.
These results illustrate the dominance of the dynamic programs to compute both upper and lower bounds.
The \ngr relaxation always finds the optimal solution, except once.
No time limit was given to this algorithm, but we observe in Table \ref{tab:empplc-time} that it actually runs very quickly.
\stsf provides poorer results with the same average running time, which, besides, is bounded by the time limit.
Similarly, the \cd is very powerful as it finds the optimal solution for 67 instances out of 81.
Even when it does not succeed to find the optimum, the gap is the smallest among every feasible solutions.
On the contrary, the local search sometimes (for 13 instances) fails to find a negative solution when there is one.

Most importantly, the \cd and the \ngr always return a solution which has the same sign as the optimal solution in a small amount of time.
Thus, these algorithms are really suitable for the pricing step of a \cg scheme.

\begin{table}[htbp]
\centering
\renewcommand{\arraystretch}{1.5}
\begin{tabular}{lcccc}
\toprule
 & CM$^{**}$ & LM$^{**}$ & TSF$^*$ &  \ngm  \\
\midrule
Average & 	1.17	&	1.05 	&	0.44		&	0.43 \\
Maximum &  	2.39	&	3.05	&	1.51	&	2.76\\
\bottomrule
\end{tabular}
 \captionsetup{justification=centering}
\caption{Average and maximum running times to get solutions\\
{ \footnotesize
$^*$ \ Time limit of 1 second \hspace*{\fill} \\
$^{**}$ Soft time limit of 1 second (checked only between iterations) \hspace*{\fill}}
\label{tab:empplc-time}}
\end{table}

\section{Solving the \kep\label{sec:CG}}

In the previous sections, algorithms and results are provided for the \empplc.
Our initial goal is however to solve the \kep.
In this section, we present experimental results comparing the performance of the previous algorithms in a \cg scheme.

\subsection{Complete algorithm}

\begin{figure}[htb]
\centering
\includegraphics[scale=1]{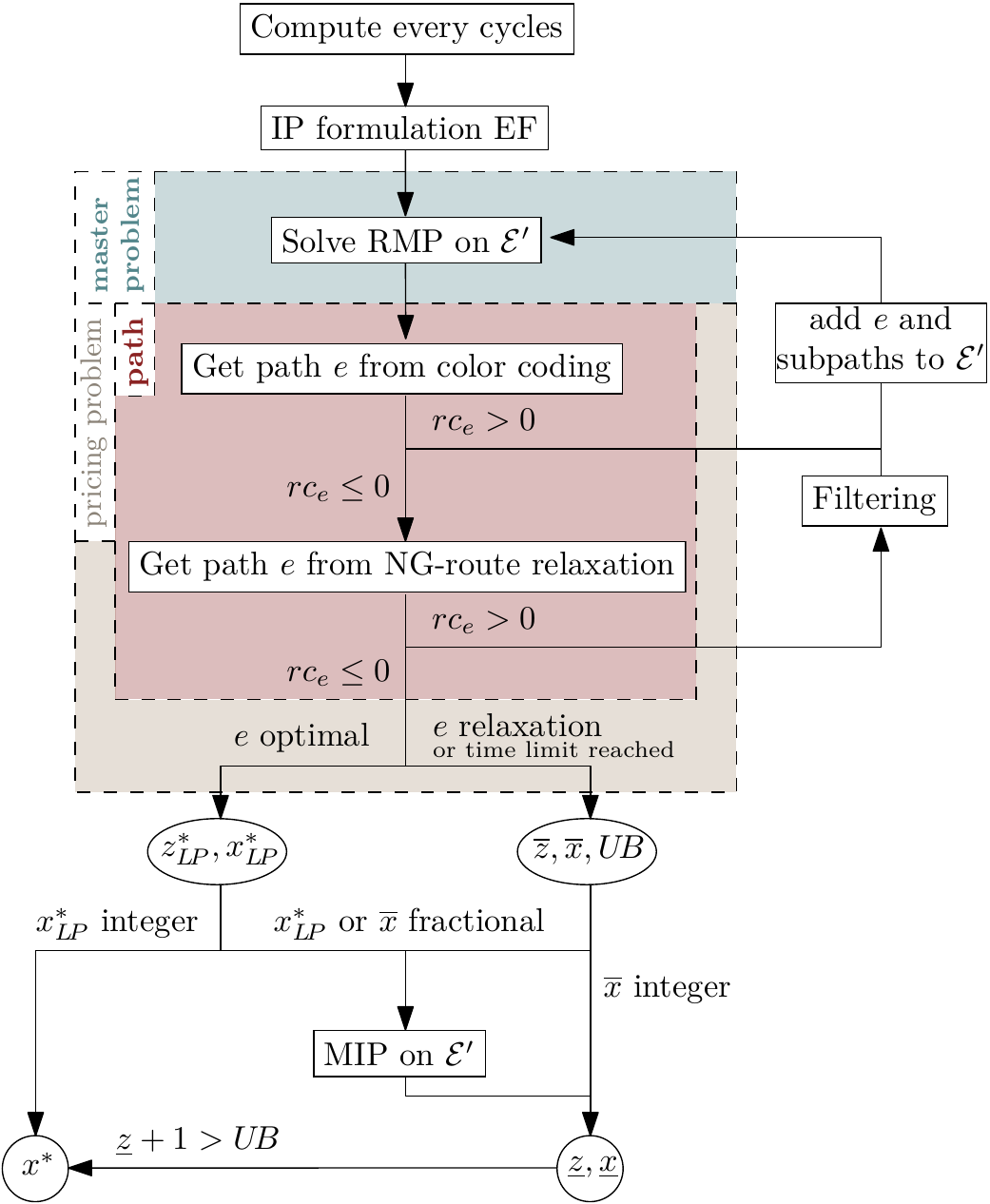}
\caption{Our algorithm solving the \kep: \cdng \label{fig:cg-real}}
\end{figure}

The complete framework of our implementation called \emphdefw{CG-dyn} is shown in Figure \ref{fig:cg-real}. 
It solves only the path pricing problem while cycles of donation are all computed beforehand and form the initial set of variables $\exchangeset'$.
This is more efficient when $\k=3$ than solving the complete pricing problem as preliminary experiments revealed.

The path pricing problem is solved with the two dynamic programs (\cd and \ngr) studied in this article.
The \cd is limited to 1 second at each iteration of the \cg. 
Then, if the \cd fails to find an improving path within this time limit, the \ngr relaxation is called.
\corbis{
Recall that the \skep is a maximization problem so a path $e$ should have a \textbf{positive} reduced cost $rc_e$ to be added in the \cg.
When the \ngr relaxation does not find an improving path, \ie every reduced cost is negative or zero, then the \cg stops.
If the solution of this last \ngr relaxation was elementary, thus optimal, then the  linear relaxation $\zexfl$ is returned. 
Otherwise, the last value of RMP $\bar{z}$ is only an upper bound $U\!B$ on this value.
If a given time limit was reached before finishing the \cg, then we cannot conclude on the link between $\zexfl$ and the current value of RMP $\bar{z}$. 
However, we can still compute an upper bound $U\!B$ with Lagrangian relaxation.
This upper bound is computed by relaxing in the objective function the redundant constraint stating that there are at most $|V| / 2 $ exchanges in the compatibility graph (see \cite{pansart2018mosim} for more details).

%
%

 }

%
%
%
%
%
After the \cg, the final integer solution \corbis{$\underline{x}$} is computed with the integer program \exf restricted on $\exchangeset'$ as it runs very quickly, especially compared to the \cg running time.
\corbis{This step is not required if the last computed solution $\bar{x}$ is already integer.}
The \ngr relaxation is also used to filter suboptimal arcs and vertices of the \kep instance.

These results also  demonstrate that adding all the subpaths of the \sempplc solution speeds up the \cg.
Concretely, when the path \boxm{(s,v_1,...,v_l)} is added to $\exchangeset'$, we also add the paths \boxm{(s,v_1,...,v_{l-1})}, \boxm{(s,v_1,...,v_{l-2})}, and so on until \boxm{(s,v_1,v_2)}.

%
%
%
%



\subsection{Performance of the pricing step \label{sec:cg-analysis}}

The efficiency of dynamic programming approaches to solve the \empplc was demonstrated in Section~\ref{sec:empplc-exp}.
To find out if this is still the case when they are embedded in a \cg scheme solving the \skep, we compared our algorithm \cdng with another \cg called \lsts, which handles the pricing step with the local search  heuristic and the \tsf.
Every other parameter of the \cg framework is the same for both algorithms, in particular they generate all cycles in the master problem, add every subpath in $\exchangeset'$, and solve a restricted \exf at the end instead of solving a complete \bp.
They are applied on 135 instances (5 in each class of \kbear)\footnote{available at \url{https://pagesperso.g-scop.grenoble-inp.fr/\~pansartl/data/instances-KBR.zip}}, each within a time limit of 2000 seconds.

\begin{table}[htb]
\captionsetup{justification=centering}
\centering
\begin{tabular}{lcc}
\toprule
& \cdng & \lsts \\
\midrule 
\# $\zexfl$ computed & 135 & 120\\
Average gap between UB and LB & 0.13 \% & 3.02 \%\\
\# $z^*$ computed (gap = 0) & 93 & 84 \\
\bottomrule
\end{tabular}
\caption{Results for the two different \scg schemes on 135 instances after 2000 seconds of running time \label{tab:cg-2cg}}
\end{table}

\begin{table}[htb]
\captionsetup{justification=centering}
\centering
\begin{tabular}{lcc}
\toprule
\l & \cdng & \lsts \\
\midrule 
4 & 1.6 & 8.9\\
7 & 10  & 34.3 \\
13 & 15.6 & 673.1 \\
\bottomrule
\end{tabular}
\caption{Average running time (in seconds), depending on $\l$, for the 120 instances solved in the  time limit\label{tab:cg-2cg-time}}
\end{table}

Table \ref{tab:cg-2cg} shows that our algorithm always finds the linear relaxation \sexfl while \lsts reaches the time limit for 15 instances (those  with \boxm{|P| = 250} and \boxm{\l = 13}).
Consequently, the \cdng algorithm finds the optimal solution of the integer program \exf for more instances.
For both algorithms, this integer solution is mostly found because the linear relaxation is integer (and valid), and in this case there is no need to call the integer program after the \cg.
Moreover, considering the 120 instances for which both methods find the linear relaxation in 2000 seconds, the running time is significantly smaller for \cdng than \lsts (see Table \ref{tab:cg-2cg-time}).
These results support the efficiency of the dynamic programs to solve the path pricing problem in a \cg for the \skep.

\subsection{Scaling up}

As \kepgs are growing, we experiment \cdng on larger instances, generated as before but with different parameters.
In particular, it seems reasonable to consider that most of the altruistic donors are already included in \kepgs, unlike patients for whom such programs are very different from the standard procedure.
Thus, the proportion of altruistic donors would probably be low in future large programs.
In the end, we apply \cdng on 20 instances\footnote{available at \url{https://pagesperso.g-scop.grenoble-inp.fr/\~pansartl/data/instances-KBL.zip}}, divided in   4 classes (\boxm{\l \in \{4,7\}}, \boxm{|P| \in \{500,750\}}, \boxm{p_{|N|} = 1\%}).

%
%

Results, summarized in Table \ref{tab:cg-large}, show that the quality of the solution is still very high.
Every instance was solved in less than half an hour and in average rather quickly.
We also tried to solve instances with \boxm{\l = 13} or \boxm{|P| = 1000},  but encountered memory issues.
However, our implementation does not profit from the fact that instances are quite sparse, so a more efficient implementation should overcome these memory errors.
\corter{Moreover, while solving the final integer program corresponds in average to less than 1\% of the running time on realistic instances, for these large instances it represents more than 5\%, sometimes almost 15\%.
It is likely that this proportion will increase with the size of instances, making necessary the development of new algorithms to find feasible solutions.}

\begin{table}[htb]
\captionsetup{justification=centering}
\centering
\begin{tabular}{lcccc}
\toprule
&  \multicolumn{2}{c}{$|P|=500$} &  \multicolumn{2}{c}{$|P|=750$}  \\
\cmidrule(lr){2-3} \cmidrule(lr){4-5}
&  $\l=4$  & $\l=7$ &  $\l=4$  & $\l=7$    \\
\midrule 
Average gap between UB and LB&  0.05\% & 0.18\% & 0.06\% & 0.23\%  \\
Average running time (seconds) & 5.6 & 123.6 & 23.1 & 1823.5  \\
\bottomrule
\end{tabular}
\caption{Results of \cdng on 40 large instances \label{tab:cg-large}}
\end{table}

\section{Conclusion}

In this article, we designed a complete \cg framework to solve the \kep including altruistic donors.
 Due to the hardness of the pricing problem in this case, an extension of previous \cg schemes containing only cycles was not possible.
We therefore  proposed a new framework showing excellent results on realistic instances and promising for larger instances.
We believe that the memory issues encountered for instances with $|P | = 1000$ can be avoided with an implementation of the column generation
using algorithms and data structures adapted to large and sparse instances.
For example, the preprocessing step computing the distances between each
pair of vertices could be performed with a Johnson’s algorithm instead of
Floyd-Warshall. 

This \cg relies on the efficiency of the dynamic approaches to handle the pricing problem, referred to as the \empplc.
This problem has to be solved many times in the \cg framework.
We adapted algorithms from the literature of shortest paths problem to better fit the specificity of our problem and designed an experimental protocol to assess their quality.
The development of a new method for the \cd algorithm,  including a preprocessing phase ordering the vertices, a new procedure to color the graph and a shifting technique, is promising for many applications.
Indeed, it guarantees to find the optimal solution in only $C$ calls to the dynamic program in some particular cases and in any case outperforms the original discrete uniform strategy.  
Our complete study of this method can be found in \cite{pansart2020ecai}.
Other improvements can be considered to get even better results.
In particular,  an adaptation of the different techniques proposed by Pecin \al~\cite{pecin2017cuts}, including  memory cuts, would probably strengthen our implementation of the \ngr relaxation.
We also intend to study the application of the \cd for other pricing problems where it is surprisingly absent, for example for vehicle routing problems.

\end{sloppypar}
\scriptsize{
\bibliographystyle{apalike}
\bibliography{EJOR}

\begin{thebibliography}{}

\bibitem[Abraham et~al., 2007]{abraham2007clearing}
Abraham, D.~J., Blum, A., and Sandholm, T. (2007).
\newblock {Clearing algorithms for barter exchange markets: Enabling nationwide
  kidney exchanges}.
\newblock In {\em Proceedings of the 8th ACM conference on Electronic
  commerce}, pages 295--304. ACM.

\bibitem[Adolphson and Hu, 1973]{adolphson1973optimal}
Adolphson, D. and Hu, T.~C. (1973).
\newblock {Optimal linear ordering}.
\newblock {\em SIAM Journal on Applied Mathematics}, 25(3):403--423.

\bibitem[Alon et~al., 1995]{alon1995color}
Alon, N., Yuster, R., and Zwick, U. (1995).
\newblock {Color-coding}.
\newblock {\em Journal of the ACM (JACM)}, 42(4):844--856.

\bibitem[Baldacci et~al., 2011]{baldacci2011ngroute}
Baldacci, R., Mingozzi, A., and Roberti, R. (2011).
\newblock {New Route Relaxation and Pricing Strategies for the Vehicle Routing
  Problem}.
\newblock {\em Operations Research}, 59(5):1269--1283.

\bibitem[Bir{\'{o}} et~al., 2017]{cost2017kidney}
Bir{\'{o}}, P., Burnapp, L., Haase, B., Hemke, A., Johnson, R., van~de
  Klundert, J., and Manlove, D. (2017).
\newblock {First handbook: Kidney exchange practices in Europe}.
\newblock Technical report, European Network for Collaboration on Kidney
  Exchange Programmes.

\bibitem[Bir{\'{o}} et~al., 2019]{biro2019building}
Bir{\'{o}}, P., Haase-Kromwijk, B., Andersson, T., {\'{A}}sgeirsson, E.~I.,
  Baltesov{\'{a}}, T., Boletis, I., Bolotinha, C., Bond, G., B{\"{o}}hmig, G.,
  Burnapp, L., and Others (2019).
\newblock {Building kidney exchange programmes in Europe—an overview of
  exchange practice and activities}.
\newblock {\em Transplantation}, 103(7):1514.

\bibitem[Biro et~al., 2009]{biro2009maximum}
Biro, P., Manlove, D.~F., and Rizzi, R. (2009).
\newblock {Maximum weight cycle packing in directed graphs, with application to
  kidney exchange programs}.
\newblock {\em Discrete Mathematics, Algorithms and Applications},
  1(04):499--517.

\bibitem[Chen et~al., 2011]{chen2011computerized}
Chen, Y., Kalbfleisch, J.~D., Li, Y., Song, P. X.~K., and Zhou, Y. (2011).
\newblock {Computerized platform for optimal organ allocations in kidney
  exchanges}.
\newblock In {\em Proceedings of the International Conference on Bioinformatics
  {\&} Computational Biology (BIOCOMP)}, page~1. The Steering Committee of The
  World Congress in Computer Science.

\bibitem[{De Klerk} et~al., 2005]{de2005dutch}
{De Klerk}, M., Keizer, K.~M., Claas, F. H.~J., Witvliet, M., Haase-Kromwijk,
  B. J. J.~M., and Weimar, W. (2005).
\newblock {The Dutch national living donor kidney exchange program}.
\newblock {\em American Journal of Transplantation}, 5(9):2302--2305.

\bibitem[Desrochers, 1988]{desrochers1988spprc}
Desrochers, M. (1988).
\newblock {An algorithm for the shortest path problem with resource
  constraints}.
\newblock Technical Report G-88-27, GERAD, Ecole des HEC, Canada.

\bibitem[Dickerson et~al., 2016]{dickerson2016position}
Dickerson, J.~P., Manlove, D.~F., Plaut, B., Sandholm, T., and Trimble, J.
  (2016).
\newblock {Position-indexed formulations for kidney exchange}.
\newblock In {\em Proceedings of the 2016 ACM Conference on Economics and
  Computation}, pages 25--42. ACM.

\bibitem[Ellison, 2014]{ellison2014systematic}
Ellison, B. (2014).
\newblock {A systematic review of kidney paired donation: Applying lessons from
  historic and contemporary case studies to improve the US model}.
\newblock Technical report, The Wharton School of the University of
  Pennsylvania.

\bibitem[Feillet et~al., 2004]{feillet2004espprc}
Feillet, D., Dejax, P., Gendreau, M., and Gueguen, C. (2004).
\newblock {An Exact Algorithm for the Elementary Shortest Path Problem with
  Resource Constraints: application to some Vehicle and Routing Problems}.
\newblock {\em Networks: An International Journal}, 44(3):216--229.

\bibitem[Fox et~al., 1980]{fox1980n}
Fox, K.~R., Gavish, B., and Graves, S.~C. (1980).
\newblock {An n-constraint formulation of the (time-dependent) traveling
  salesman problem}.
\newblock {\em Operations Research}, 28(4):1018--1021.

\bibitem[Garey et~al., 1974]{garey1974some}
Garey, M.~R., Johnson, D.~S., and Stockmeyer, L. (1974).
\newblock {Some simplified NP-complete problems}.
\newblock In {\em Proceedings of the sixth annual ACM symposium on Theory of
  computing}, pages 47--63. ACM.

\bibitem[Glorie et~al., 2012]{glorie2012iterative}
Glorie, K., Wagelmans, A., and van~de Klundert, J. (2012).
\newblock {Iterative branch-and-price for large multi-criteria kidney
  exchange}.
\newblock Technical report, Econometric institute, Erasmus University
  Rotterdam.

\bibitem[Glorie et~al., 2014]{glorie2014kidney}
Glorie, K.~M., van~de Klundert, J.~J., and Wagelmans, A. P.~M. (2014).
\newblock {Kidney exchange with long chains: An efficient pricing algorithm for
  clearing barter exchanges with branch-and-price}.
\newblock {\em Manufacturing {\&} Service Operations Management},
  16(4):498--512.

\bibitem[Held and Karp, 1962]{held1962dynamic}
Held, M. and Karp, R.~M. (1962).
\newblock {A dynamic programming approach to sequencing problems}.
\newblock {\em Journal of the Society for Industrial and Applied Mathematics},
  10(1):196--210.

\bibitem[H{\"{u}}ffner et~al., 2008]{huffner2008algorithm}
H{\"{u}}ffner, F., Wernicke, S., and Zichner, T. (2008).
\newblock {Algorithm engineering for color-coding with applications to
  signaling pathway detection}.
\newblock {\em Algorithmica}, 52(2):114--132.

\bibitem[Klimentova et~al., 2014]{klimentova2014new}
Klimentova, X., Alvelos, F., and Viana, A. (2014).
\newblock {A new branch-and-price approach for the kidney exchange problem}.
\newblock In {\em International Conference on Computational Science and Its
  Applications}, pages 237--252. Springer.

\bibitem[Kneis et~al., 2011]{kneis2009derandomizing}
Kneis, J., Langer, A., and Rossmanith, P. (2011).
\newblock {Derandomizing Non-uniform Color-Coding I}.
\newblock Technical report, RWTH Aachen - Department of Computer Science.

\bibitem[Kramer et~al., 2019]{kramer2019european}
Kramer, A., Pippias, M., Noordzij, M., Stel, V.~S., Andrusev, A.~M.,
  Aparicio-Madre, M.~I., {Arribas Monz{\'{o}}n}, F.~E., {\AA}sberg, A.,
  Barbullushi, M., Beltr{\'{a}}n, P., and Others (2019).
\newblock {The European Renal Association--European Dialysis and Transplant
  Association (ERA-EDTA) Registry Annual Report 2016: a summary}.
\newblock {\em Clinical kidney journal}, 12(5):702--720.

\bibitem[Kwak et~al., 1999]{kwak1999exchange}
Kwak, J.~Y., Kwon, O.~J., Lee, K.~S., Kang, C.~M., Park, H.~Y., and Kim, J.~H.
  (1999).
\newblock {Exchange-donor program in renal transplantation: a single-center
  experience}.
\newblock {\em Transplantation proceedings}, 31(1):344--345.

\bibitem[Mak-Hau, 2017]{mak2017kidney}
Mak-Hau, V. (2017).
\newblock {On the kidney exchange problem: cardinality constrained cycle and
  chain problems on directed graphs: a survey of integer programming
  approaches}.
\newblock {\em Journal of Combinatorial Optimization}, 33(1):35--59.

\bibitem[Pansart et~al., 2020]{pansart2020ecai}
Pansart, L., Cambazard, H., and Catusse, N. (2020).
\newblock {New randomized strategies for the color coding algorithm}.
\newblock In {\em ECAI 2020}. IOS Press.

\bibitem[Pansart et~al., 2018]{pansart2018mosim}
Pansart, L., Cambazard, H., and Stauffer, N. C.~G. (2018).
\newblock {Column Generation for the Kidney Exchange Problem}.
\newblock In {\em 12th International Conference on MOdeling, Optimization and
  SIMulation - MOSIM18 - June 27-29 2018 Toulouse - France "The rise of
  connected systems in industry and services"}.

\bibitem[Park et~al., 2004]{park2004relay}
Park, J.-H., Park, J.-W., Koo, Y.-M., and Kim, J.~H. (2004).
\newblock {Relay kidney transplantation in Korea—legal, ethical and medical
  aspects}.
\newblock {\em Legal Medicine}, 6(3):178--181.

\bibitem[Pecin et~al., 2017]{pecin2017cuts}
Pecin, D., Pessoa, A., Poggi, M., Uchoa, E., and Santos, H. (2017).
\newblock {Limited memory rank-1 cuts for vehicle routing problems}.
\newblock {\em Operations Research Letters}, 45(3):206--209.

\bibitem[Pecin et~al., 2013]{pecin2013ngroute}
Pecin, D., Poggi, M., and Martinelli, R. (2013).
\newblock {Efficient elementary and restricted non-elementary route pricing}.
\newblock Technical report, Pontifical Catholic University of Rio de Janeiro.

\bibitem[Plaut et~al., 2016a]{plaut2016fast}
Plaut, B., Dickerson, J.~P., and Sandholm, T. (2016a).
\newblock {Fast optimal clearing of capped-chain barter exchanges}.
\newblock In {\em AAAI Conference on Artificial Intelligence (AAAI)}.

\bibitem[Plaut et~al., 2016b]{plaut2016hardness}
Plaut, B., Dickerson, J.~P., and Sandholm, T. (2016b).
\newblock {Hardness of the pricing problem for chains in barter exchanges}.

\bibitem[Rapaport, 1986]{rapaport1986case}
Rapaport, F.~T. (1986).
\newblock {The case for a living emotionally related international kidney donor
  exchange registry.}
\newblock {\em Transplantation proceedings}, 18(3 Suppl. 2):5--9.

\bibitem[Righini and Salani, 2008]{righini2008new}
Righini, G. and Salani, M. (2008).
\newblock {New Dynamic Programming Algorithms for the Resource Constrained
  Elementary Shortest Path Problem}.
\newblock {\em Networks: An International Journal}, 51(3):155--170.

\bibitem[Saidman et~al., 2006]{saidman2006increasing}
Saidman, S.~L., Roth, A.~E., S{\"{o}}nmez, T., {\"{U}}nver, M.~U., and
  Delmonico, F.~L. (2006).
\newblock {Increasing the opportunity of live kidney donation by matching for
  two-and three-way exchanges}.
\newblock {\em Transplantation}, 81(5):773--782.

\bibitem[Saran et~al., 2017]{saran2017us}
Saran, R., Robinson, B., Abbott, K., Agodoa, L., Albertus, P., Ayanian, J.,
  Balkrishnan, R., Bragg-Gresham, J., Cao, J., Chen, J., and Others (2017).
\newblock {US renal data system 2016 annual data report: epidemiology of kidney
  disease in the United States}.
\newblock {\em American journal of kidney diseases: the official journal of the
  National Kidney Foundation}, 69(3):A7--A8.

\bibitem[Saran et~al., 2018]{saran2018us}
Saran, R., Robinson, B., Abbott, K.~C., Agodoa, L. Y.~C., Bhave, N.,
  Bragg-Gresham, J., Balkrishnan, R., Dietrich, X., Eckard, A., Eggers, P.~W.,
  and Others (2018).
\newblock {US renal data system 2017 annual data report: epidemiology of kidney
  disease in the United States}.
\newblock {\em American journal of kidney diseases: the official journal of the
  National Kidney Foundation}, 71(3 Suppl 1):A7.

\end{thebibliography}
  }

\end{document}